\documentclass[journal,12pt, draftcls, letterpaper,onecolumn]{IEEEtran}

\usepackage{amsfonts}
\usepackage{amsmath}
\usepackage{amssymb}

\usepackage{verbatim}
\usepackage{graphicx}

\newtheorem{theorem}{Theorem}
\newtheorem{definition}{Definition}

\newtheorem{proposition}{Proposition}

\newtheorem{conjecture}{Conjecture}

\DeclareMathOperator{\p}{\mathrm{p}}
\DeclareMathOperator*{\pspace}{(\Omega, \mathcal{F}, \mathbf{P})}
\DeclareMathOperator*{\state}{S}

\DeclareMathOperator*{\sspace}{(\state,\mathcal{S})}
\DeclareMathOperator{\meet}{\wedge}
\DeclareMathOperator{\join}{\vee}

\DeclareMathOperator*{\conn}{\overset{\scriptscriptstyle G }{\sim}}
\DeclareMathOperator{\lattice}{\mathrm{L}}
\providecommand{\gen}[1]{\langle#1\rangle}

\begin{document}

\title{A Group Theoretic Model for Information}

\author{Hua~Li and Edwin K.~P.~Chong,~\IEEEmembership{Fellow,~IEEE}%
\thanks{A preliminary version of the material in this paper was presented in part at the 45th Annual Allerton Conference on Communication, Control, and Computing, Monticello, Illinois, September 26--28, 2007.}%
\thanks{H.~Li is with the Department of Electrical and Computer Engineering, Colorado State University, Fort Collins, CO 80523 (email: Hua.Li@ColoState.Edu).}%
\thanks{E.~K.~P.~Chong is with the Department of Electrical and Computer Engineering and the Department of Mathematics, Colorado State University, Fort Collins, CO 80523 (email: Edwin.Chong@ColoState.Edu).}%
}

\maketitle

\begin{abstract}
In this paper we formalize the notions of information elements and information lattices, first proposed by Shannon. Exploiting this formalization, we identify a comprehensive parallelism between information lattices and subgroup lattices. Qualitatively, we demonstrate isomorphisms between information lattices and subgroup lattices. Quantitatively, we establish a decisive approximation relation between the entropy structures of information lattices and the log-index structures of the corresponding subgroup lattices. This approximation extends the approximation for joint entropies carried out previously by Chan and Yeung. As a consequence of our approximation result, we show that any continuous law holds in general for the entropies of information elements if and only if the same law holds in general for the log-indices of subgroups. As an application, by constructing subgroup counterexamples we find surprisingly that common information, unlike joint information, obeys neither the submodularity nor the supermodularity law.  We emphasize that the notion of information elements is conceptually significant---formalizing it helps to reveal the deep connection between information theory and group theory. The parallelism established in this paper admits an appealing group-action explanation and provides useful insights into the intrinsic structure among information elements from a group-theoretic perspective.
\end{abstract}


\section{Introduction}
Information theory was born with the celebrated entropy formula measuring the \emph{amount} of information for the purpose of communication. However, a suitable mathematical model for \emph{information itself} remained elusive over the last sixty years. It is reasonable to assume that information theorists have had certain intuitive conceptions 
of information, but in this paper we seek a mathematic model for such a conception. In particular, building on Shannon's work~\cite{Shannon53:1}, we formalize the notion of \emph{information elements} to capture the syntactical essence of \emph{information}, and identify information elements with $\sigma$-algebras and \emph{sample-space-partitions}. As we shall see in the following, by building such a mathematical model for information and identifying the lattice structure among information elements, the seemingly surprising connection between information theory and group theory, established by Chan and Yeung~\cite{Chan02}, is revealed via isomorphism relations between information lattices and subgroup lattices. Consequently, a fully-fledged and decisive approximation relation between the entropy structure of information lattices and the subgroup-index structure of corresponding subgroup lattices is obtained.

We first motivate our formal definition for the notion of information elements.
\subsection{Informationally Equivalent Random Variables}
Recall the profound insight offered by Shannon~\cite{Shannon48} on the essence of communication: ``the fundamental problem of communication is that of reproducing at one point exactly or approximately a message selected at another point.'' Consider the following motivating example. Suppose a message, \emph{in English}, is delivered from person A to person B. Then, the message is translated and delivered \emph{in German} by person B to person C (perhaps because person C does not know English). Assuming the translation is faithful, person C should receive the message that person A intends to convey. Reflecting upon this example, we see that the message (information) assumes two different ``representations'' over the process of the entire communication---one in English and the other in German, but the message (information) itself remains the same. Similarly, coders (decoders), essential components of communication systems, perform the similar function of ``translating'' one representation of the same information to another one. This suggests that ``information'' itself should be defined in a translation invariant way. This ``translation-invariant'' quality is precisely how we seek to characterize information.

To introduce our formal definition for information elements to capture the essence of information itself, we note that information theory is built within the probabilistic framework, in which one-time information sources are usually modeled by random variables. Therefore, we start in the following with the concept of \emph{informational equivalence} between random variables and develop the formal concept of information elements from first principles.

Recall that, given a probability space $\pspace$ and a measurable space $\sspace$, a random variable is a measurable function from $\Omega$ to $\state$. The set $\state$ is usually called the state space of the random variable, and $\mathcal{S}$ is a $\sigma$-algebra on $\state$. The set $\Omega$ is usually called the \emph{sample space}; $\mathcal{F}$ is a $\sigma$-algebra on $\Omega$, usually called the \emph{event space}; and  $\mathbf{P}$ denotes a probability measure on the measurable space $(\Omega, \mathcal{F})$.

To illustrate the idea of \emph{informational equivalence}, consider a random variable $X:\Omega \rightarrow \state$ and another random variable $X'=f(X)$, where the function $f: \state \rightarrow \state'$ is bijective. Certainly, the two random variables $X$ and $X'$ are \emph{technically different} for they have different codomains. However, it is intuitively clear that that they are ``equivalent'' in some sense. In particular, one can infer the exact state of $X$ by observing that of $X'$, and vice versa. For this reason, we may say that the two random variables $X$ and $X'$ carry the same piece of information. Note that the $\sigma$-algebras induced by $X$ and $X'$ coincide with each other. In fact, two random variables such that the state of one can be inferred from that of the other induce the same $\sigma$-algebra. This leads to the following definition for \emph{information equivalence}.

\begin{definition}
We say that two random variables $X$ and $X'$ are \emph{informationally equivalent}, denoted $X\cong X'$, if the $\sigma$-algebras induced by $X$ and $X'$ coincide.
\end{definition}

It is easy to verify that the ``being-informational-equivalent'' relation is an equivalence relation. The definition reflects our intuition, as demonstrate in the previous motivating examples, that two random variables carry the same piece information if and only if they induce the same $\sigma$-algebra. This motivates the following definition for \emph{information elements} to capture the syntactical essence of information itself.

\begin{definition}
An \emph{information element} is an equivalence class of random variables with respect to the ``being-informationally-equivalent'' relation.
\end{definition}

We call the random variables in the equivalent class of an information element $m$ \emph{representing random variables} of $m$. Or, we say that a random variable $X$ \emph{represents} $m$.

We believe that our definition of information elements reflects exactly Shannon's original intention~\cite{Shannon53:1}:

\begin{quote}
Thus we are led to define the actual information of a stochastic process as that which is common to all stochastic processes which may be obtained from the original by reversible encoding operations.
\end{quote}

Intuitive (also informal) discussion on identifying ``information'' with $\sigma$-algebras surfaces often in probability theory, martingale theory, and mathematical finance. In probability theory, see for example~\cite{Billingsley95}, the concept of conditional probability is usually introduced with discussion of treating the $\sigma$-algebras conditioned on as the ``partial information'' available to ``observers.'' In martingale theory and mathematical finance, see for example~\cite{Shreve05,Ankirchner06}, \emph{filtrations}---increasing sequences of $\sigma$-algebras---are often interpreted as records of the information available over time.

\subsubsection{A Few Observations}
\begin{proposition}\label{equivalence.entropy}
If $X\cong X'$, then $H(X)=H(X')$.
\end{proposition}

(Throughout the paper, we use $H(X)$ to denote the entropy of random variable $X$.)

The conserve to Proposition~\ref{equivalence.entropy} fails---two random variables with a same entropy do not necessarily carry the same information. For example, consider two binary random variables $X,Y: \Omega \rightarrow \{0,1\}$, where $\Omega=\{a,b,c,d\}$ and $\mathbf{P}$ is uniform on $\Omega$. Suppose $X(\omega)=0$ if $\omega=a, b$ and 1 otherwise, and $Y(\omega)=0$ if $\omega=a,c$ and 1 otherwise. Clearly, we have $H(X)=H(Y)=1$, but one can readily agree that $X$ and $Y$ do \emph{not} carry the same information. Therefore, the notion of ``informationally-equivalent'' is stronger than that of ``identically-distributed.''

On the other hand, we see that the notion of ``informationally-equivalent'' is weaker than that of ``being-equal.''
\begin{proposition}\label{almost.surely}
If $X=X'$, then $X\cong X'$.
\end{proposition}

The converse to Proposition~\ref{almost.surely} fails as well, since two informationally equivalent random variable $X$ and $X'$ may have totally different state spaces, so that it does not even make sense to say $X=X'$.

As shown in the following proposition, the notion of ``informational equivalence'' characterizes a kind of state space invariant ``equalness.''
\begin{proposition}
Two random variables $X$ and $Y$ with state spaces $\mathcal{X}$ and $\mathcal{Y}$, respectively, are informationally equivalent if and only if there exists a one-to-one correspondence $f: \mathcal{X} \rightarrow \mathcal{Y}$ such that $Y=f(X)$.
\end{proposition}

Remark: Throughout the paper, we fix a probability space unless otherwise stated. For ease of presentation, we confine ourselves in the following to finite discrete random variables. However, most of the definitions and results can be applied to more general settings without significant difficulties.

\subsection{Identifying Information Elements via $\sigma$-algebras and Sample-Space-Partitions}
Since the $\sigma$-algebras induced by informationally equivalent random variables are the same, we can unambiguously identify information elements with $\sigma$-algebras. Moreover, because we deal with finite discrete random variables exclusively in this paper, we can afford to discuss $\sigma$-algebras more explicitly as follows.

Recall that a \emph{partition} $\Pi$ of a set $A$ is a collection $\{\pi_i:i\in[k] \}$ of disjoint subsets of $A$ such that $\cup_{i\in[k]} \pi_i= A$. (Throughout the paper, we use the bracket notation $[k]$ to denote the generic index set $\{1,2,\cdots,k\}$.) The elements of a partition $\Pi$ are usually called the \emph{parts} of $\Pi$. It is well known that there is a natural one-to-one correspondence between partitions of the sample space and the $\sigma$-algebras---any given $\sigma$-algebra of a sample space can be generated uniquely, via union operation, from the atomic events of the $\sigma$-algebra, while the collection of the atomic events forms a partition of the sample space. For example, for a random variable $X: \Omega \rightarrow \mathcal{X}$, the atomic events of the $\sigma$-algebra induced by $X$ are $X^{-1}(\{x\}), x\in \mathcal{X}$. For this reason, from now on, we shall identify an information element by either its $\sigma$-algebra or its corresponding sample space partition.

It is well known that the number of distinct partitions of a set of size $n$ is the $n$th Bell number and that the Stirling number of the second kind $S(n,k)$ counts the number of ways to partition a set of $n$ elements into $k$ nonempty parts. These two numbers, crucial to the remarkable results obtained by Orlitsky et al. in~\cite{Orlitsky04}, suggest a possibly interesting connection between the notion of information elements discussed in this paper and the ``patterns'' studied in~\cite{Orlitsky04}.
\subsection{Shannon's Legacy}
As we mentioned before, the notion of \emph{information elements} was originally proposed by Shannon in \cite{Shannon53:1}. In the same paper, Shannon also proposed a partial order for information elements and a lattice structure for collections of information elements. We follow Shannon and call such lattices \emph{information lattices} in the following.

Abstracting the notion of information elements out of their representations---random variables---is a conceptual leap, analogous to the leap from the concrete calculation with matrices to the study of abstract vector spaces. To this end, we formalize both the ideas of information elements and information lattices. By identifying information elements with sample-space-partitions, we are equipped to establish a comprehensive parallelism between information lattices and subgroup lattices. Qualitatively, we demonstrate isomorphisms between information lattices and certain subgroup lattices. With such isomorphisms established, quantitatively, we establish an approximation for the entropy structure of information lattices, consisting of joint, common, and many other information elements, using the log-index structures of their counterpart subgroup lattices. Our approximation subsumes the approximation carried out only for joint information elements by Chan and Yeung~\cite{Chan02}. Building on~\cite{Chan02}, the parallelism identified in this paper reveals an intimate connection between information theory and group theory and suggests that group theory may provide suitable mathematical language to describe and study laws of information.

The full-fledged parallelism between information lattices and subgroup lattices established in paper is one of our main contributions. With this intrinsic mathematical structure among multiple information elements being uncovered, we anticipate more systematic attacks on certain network information problems, where a better understanding of intricate internal structures among multiple information elements is in urgent need. Indeed, the ideas of information elements and information lattices were originally motivated by network communication problems---in~\cite{Shannon53:1}, Shannon wrote:
\begin{quote}
The present note outlines a new approach to information theory which is aimed specifically at the analysis of certain communication problems in which there exist a number of sources simultaneously in operation.
\end{quote}
and
\begin{quote}
Another more general problem is that of a communication system consisting of a large number of transmitting and receiving points with some type of interconnecting network between the various points. The problem here is to formulate the best system design whereby, in some sense, the best overall use of the available facilities is made.
\end{quote}
It is not hard to see that Shannon was attempting to solve now-well-known network coding capacity problems.

Certainly, we do not claim that all the ideas in this paper are our own. For example, as we pointed out previously, the notions of information elements and information lattices were proposed as early as the 1950s by Shannon~\cite{Shannon53:1}. However, this paper of Shannon's is not well recognized, perhaps owing to the abstruseness of the ideas. Formalizing these ideas and connecting them to current research is one of the primary goals of this paper. For all other results and ideas that have been previously published, we separate them from those of our own by giving detailed references to their original sources.

\subsection{Organization}
The paper is organized as follows. In Section~\ref{lattice.intro}, we introduce a ``being-richer-than'' partial order between information elements and study the information lattices induced by this partial order. In Section~\ref{isomorphism}, we formally establish isomorphisms between information lattices and subgroup lattices. Section~\ref{approximation} is devoted to the quantitative aspects of information lattices. We show that the entropy structure of information lattices can be approximated by the log-index structure of their corresponding subgroup lattices. As a consequence of this approximation result, in Section~\ref{law.parallelism}, we show  that any continuous law holds for the entropies of common and joint information if and only if the same law holds for the log-indices of subgroups. As an application of this result, we show a result, which is rather surprising, that unlike joint information neither the submodularity nor the supermodularity law holds for common information in general. We conclude the paper with a discussion in Section~\ref{discussion}.

\section{Information Lattices}\label{lattice.intro}
\subsection{``Being-richer-than'' Partial Order}
Recall that every information element can be identified with its corresponding sample-space-partition. Consider two sample-space-partitions $\Pi$ and $\Pi'$. We say that $\Pi$ is \emph{finer than} $\Pi'$, or $\Pi'$ is \emph{coarser than} $\Pi$, if each part of $\Pi$ is contained in some part of $\Pi'$.
\begin{definition}
For two information elements $m_1$ and $m_2$, we say that $m_1$ is \emph{richer than} $m_2$, or $m_2$ is \emph{poorer than} $m_2$, if the sample-space-partition of $m_1$ is finer than that of $m_2$. In this case, we write $m_1\geq m_2$.
\end{definition}

It is easy to verify that the above defined ``being-richer-than'' relation is a partial order.

We have the following immediate observations:
\begin{proposition}
$m_1 \geq m_2$ if and only if $H(m_2|m_1)=0$.
\end{proposition}

As a corollary to the above proposition, we have
\begin{proposition}\label{entropy.order}
If $m_1 \geq m_2$, then $H(m_1)\geq H(m_2)$.
\end{proposition}
The converse of Proposition~\ref{entropy.order} does not hold in general.

With respect to representative random variables of information elements, we have
\begin{proposition}\label{functional}
Suppose random variables $X_1$ and $X_2$ represent information elements $m_1$ and $m_2$ respectively. Then, $m_1\geq m_2$ if and only if $X_2=f (X_1)$ for some function $f$.
\end{proposition}

A similar result to Proposition~\ref{functional} was previously observed by Renyi~\cite{Renyi70} as well.

The ``being-richer-than'' relation is very important to information theory, because it characterizes the only universal information-theoretic constraint put on all deterministic coders (decoders)---the input information element of any coder is always richer than the output information element. For example, partially via this principle, Yan et al.~recently characterized the capacity region of general acyclic multi-source multi-sink networks~\cite{Yan07}. Harvey et al.~\cite{Harvey06} obtained an improved computable outer bound for general network coding capacity regions by applying this same principle under a different name called \emph{information dominance}---the authors of the paper acknowledged: ``...information dominance plays a key role in our investigation of network capacity.''

\subsection{Information Lattices}
Recall that a lattice is a set endowed with a partial order in which any two elements have a unique supremum and a unique infimum with respect to the partial order. Conventionally, the supremum of two lattice elements $x$ and $y$ is also called the \emph{join} of $x$ and $y$; the infimum is also called the \emph{meet}. In our case, with respect to the ``being-richer-than'' partial order, the supremum of two information elements $m_1$ and $m_2$, denoted $m_1\join m_2$, is the poorest among all the information elements that are richer than both $m_1$ and $m_2$. Conversely, the infimum of $m_1$ and $m_2$, denoted $m_1\meet m_2$, is the richest among all the information elements that are poorer than both $m_1$ and $m_2$. In the following, we also use $m^{12}$ to denote the join of $m_1$ and $m_2$, and $m_{12}$ the meet.

\begin{definition}
An \emph{information lattice} is a set of information elements that is closed under the join $\join$ and meet $\meet$ operations.
\end{definition}

Recall the one-to-one correspondence between information elements and sample-space-partitions. Consequently, each information lattice corresponds to a partition lattice (with respect to the ``being-finer-than'' partial order on partitions), and vice versa. This formally confirms the assertions made in \cite{Shannon53:1}: ``they (information lattices) are at least as general as the class of finite partition lattices.''

Since the collection of information lattices could be as general as that of partition lattices, we should not expect any special lattice properties to hold generally for all information lattices, because it is well-known that any finite lattice can be embedded in a finite partition lattice~\cite{Pudlak80}. Therefore, it is not surprising to learn that information lattices are in general not distributive, not even modular.

\subsection{Joint Information Element}
The \emph{join} of two information elements is straightforward. Consider two information elements $m_1$ and $m_2$ represented respectively by two random variables $X_1$ and $X_2$. It is easy to check that the joint random variable $(X_1, X_2)$ represents the join $m^{12}$. For this reason, we also call $m^{12}$ (or $m_1\join m_2$) the \emph{joint information element} of $m_1$ and $m_2$. It is worth pointing out that the joint random variable $(X_2,X_1)$ represents $m^{12}$ equally well.

\subsection{Common Information Element}\label{characterizing.common.info}
In~\cite{Shannon53:1}, the meet of two information elements is called \emph{common information}. More than twenties years later, the same notion of common information was independently proposed and first studied in detail by G\'{a}cs and K\"{o}rner~\cite{Gacs73}. For the first time, it was demonstrated that common information could be far less than mutual information. (``Mutual information'' is rather a misnomer because it does not correspond naturally to any information element~\cite{Gacs73}.) Unlike the case of joint information elements, characterizing common information element via their representing random variables is much more complicated. See~\cite{Gacs73,Witsenhausen75} for details.

In contrast to the all-familiar joint information, common information receives far less attention. Nonetheless, it has been shown to be important to cryptography~\cite{Ahlswede93,Ahlswede98,Csiszar00,StefanWolf04}, indispensable for characterizing of the capacity region of multi-access channels with correlated sources~\cite{Cover80}, useful in studying information inequalities~\cite{Zhang03, Hammer00}, and relevant to network coding problems~\cite{Niesen05}.

\subsection{Previously Studied Lattices in Information Theory}
Historically, at least three other lattices~\cite{Fujishige78,Cicalese02,Muchnik02} have been considered in attempts to characterize certain ordering relations between information elements. Two of them, studied respectively in~\cite{Fujishige78} and~\cite{Muchnik02}, are subsumed by the information lattices considered in this paper.

\section{Isomorphisms between Information Lattices and Subgroup Lattices}\label{isomorphism}
In this section, we discuss the qualitative aspects of the parallelism between information lattices generated from sets of information elements and subgroup lattices generated from sets of subgroups. In particularly, we establish isomorphism relations between them.

\subsection{Information Lattices Generated by Information Element Sets}
It is easy to verify that both the binary operations ``$\join$'' and ``$\meet$'' are \emph{associative} and \emph{commutative}. Thus, we can readily extend them to cases of more than two information elements. Accordingly, for a given set $\{m_i: i\in [n]\}$ of information elements, we denote the joint information element of the subset $\{m_i: i\in \alpha\}$, $\alpha\subseteq [n]$, of information elements by $m^{\alpha}$ and the common information element by $m_{\alpha}$.

\begin{definition}
Given a set $\mathbf{M}=\{m_i:i\in[n]\}$ of information elements, the \emph{information lattice generated by $\mathbf{M}$}, denoted $\lattice_\mathbf{M}$, is the smallest information lattice that contains $\mathbf{M}$. We call $\mathbf{M}$ the \emph{generating set} of the lattice $\lattice_\mathbf{M}$.
\end{definition}

\begin{figure}[htp]
\centering
\includegraphics[width=0.4\textwidth]{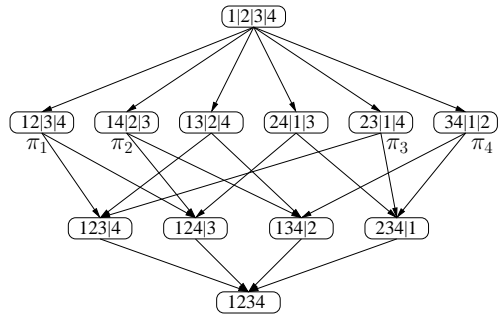}
\caption{Lattice Generated by  $\{\pi_i:i=[4]\}$ }\label{fig:Hasse}
\end{figure}

It is easy to see that each information element in $\lattice_\mathbf{M}$ can be obtained from the information elements in the generating set $\mathbf{M}$ via a sequence of join and meet operations. Note that the set $\{m_\alpha:\alpha\subseteq [n]\}$ of information elements forms a meet semi-lattice and the set $\{m^\beta:\beta\subseteq [n]\}$ forms a join semi-lattice. However, the union $\{m_\alpha,m^\beta: \alpha, \beta \subseteq [n]\}$ of these two semi-lattices does \emph{not} necessarily form a lattice. To see this, consider the following example constructed with partitions (since partitions are in one-to-one correspondence with information elements). Let $\{\pi_i:i=[4]\}$ be a collection of partitions on the set $\{1,2,3,4\}$ where $\pi_1=12|3|4$, $\pi_2=14|2|3$, $\pi_3=23|1|4$, and $\pi_4=34|1|2$. See Figure~\ref{fig:Hasse} for the Hasse diagram of the lattice generated by the collection $\{\pi_i:i=[4]\}$. It is easy to see $(\pi_1 \join \pi_2) \meet (\pi_3\join \pi_4) = 124|3 \meet 234|1=24|1|3$, but $24|1|3\notin \{\pi_\alpha,\pi^\beta:\alpha,\beta\in [4]\}$. Similarly, we have $(\pi_1 \join \pi_3) \meet (\pi_2\join \pi_4) =13|2|4 \notin \{\pi_\alpha,\pi^\beta:\alpha,\beta\in [4]\}$.

\subsection{Subgroup Lattices}
Consider the binary operations on subgroups---intersection and union. We know that the intersection $G_1 \cap G_2$ of two subgroups is again a subgroup. However, the union $G_1\cup G_2$ does \emph{not} necessarily form a subgroup. Therefore, we consider the subgroup \emph{generated} from the union $G_1 \cup G_2$, denoted $G^{12}$ (or $G_1\join G_2$). Similar to the case of information elements, the intersection and ``$\join$'' operations on subgroups are both associative and commutative. Therefore, we readily extend the two operations to the cases with more than two subgroups and, accordingly, denote the intersection $\cap_{i \in [n]} G_i$ of a set of subgroups $\{G_i:i\in [n]\}$ by $G_{[n]}$ and the subgroup generated from the union by $G^{[n]}$. It is easy to verify that the subgroups $G_{[n]}$ and $G^{[n]}$ are the infimum and the supremum of the set $\{G_i:i\in [n]\}$ with respect to the ``being-a-subgroup-of'' partial order. For notation consistency, we also use ``$\meet$'' to denote the intersection operation.

Note that, to keep the notation simple, we ``overload'' the symbols ``$\join$'' and ``$\meet$'' for both the join and the meet operations with information elements and the intersection and the ``union-generating'' operations with subgroups. Their actual meaning should be clear within context.

\begin{definition}
A \emph{subgroup lattice} is a set of subgroups that is closed under the $\meet$ and $\join$ operations.
\end{definition}

For example, the set of all the subgroups of a group forms a lattice.

Similar to the case of information lattices generated by sets of information elements, we consider in the following \emph{subgroup lattices generated by a set of subgroups}.
\begin{definition}
Given a set $\mathbf{G}=\{G_i:i\in [n]\}$ of subgroups, the \emph{subgroup lattice generated by $\mathbf{G}$}, denoted $\lattice_\mathbf{G}$, is the smallest lattices that contains $\mathbf{G}$. We call $\mathbf{G}$ the \emph{generating set} of $\lattice_{\mathbf{G}}$.
\end{definition}

Note that the set $\{G_{\alpha}:\alpha\subseteq [n]\}$ forms a semilattice under the meet $\meet$ operation and the set $\{G^\beta:\beta\subseteq [n]\}$ forms a semilattice under the join $\join$ operation. However, as in the case of information lattices, the union $\{G_{\alpha}, G^\beta: \alpha,\beta\subseteq [n]\}$ of the two semilattices does \emph{not} necessarily form a lattice.

In the remainder of this section, we relate information lattices generated by sets of information elements and subgroup lattices generated by collections of subgroups and  demonstrate isomorphism relations between them. For ease of presentation, as a special case we first introduce an isomorphism between information lattices generated by sets of coset-partition information elements and their corresponding subgroup lattices.

\subsection{Special Isomorphism Theorem}\label{coset.subsection}
We endow the sample space with a \emph{group} structure---the sample space in question is taken to be a group $G$. For any subgroup of $G$, by Lagarange's theorem~\cite{Dummit99}, the collection of its cosets forms a partition of $G$. Certainly, the coset-partition, as a sample-space-partition, uniquely defines an information element. A collection $\mathbf{G}=\{G_i:i\in [n]\}$ of subgroups of $G$, in the same spirit, identifies a set $\mathbf{M}=\{m_i:i\in [n]\}$ of information elements via this subgroup--coset-partition correspondence.

Remark: throughout the paper, groups are taken to be multiplicative, and cosets are taken to be right cosets.

It is clear that, by our construction, the information elements in $\mathbf{M}$ and the subgroups in $\mathbf{G}$ are in one-to-one correspondence via the subgroup--coset-partition relation. It turns out that the information elements on the entire information lattice $\lattice_\mathbf{M}$ and the subgroups on the subgroup lattice $\lattice_\mathbf{G}$ are in one-to-one correspondence as well via the same subgroup--coset-partition relation. In other words, both the join and meet operations on information lattices are faithfully ``mirrored'' by the join and meet operations on subgroup lattices.

\begin{theorem}{(Special Isomorphism Theorem)}\label{coset.isomorphism}
Given a set $\mathbf{G}=\{G_i:i\in [n]\}$ of subgroups, the subgroup lattice $\lattice_\mathbf{G}$ is isomorphic to the information lattice $\lattice_\mathbf{M}$ generated by the set $\mathbf{M}=\{m_i:i\in [n]\}$ of information elements, where $m_i$, $i\in [n]$, are accordingly identified via the coset-partitions of the subgroups $G_i$, $i\in [n]$.
\end{theorem}

The theorem is shown by demonstrating a mapping, from the subgroup lattice $\lattice_{\mathbf{G}}$ to the information lattice $\lattice_{\mathbf{M}}$, such that it is a lattice-morphism, i.e., it honors both join and meet operations, and is bijective as well. Naturally, the mapping $\phi: \lattice_\mathbf{G} \rightarrow \lattice_\mathbf{M}$ assigning to each subgroup $G_i\in \lattice_\mathbf{G}$ the information element identified by the coset-partition of the subgroup $G_i$ is such a morphism. Since this theorem and its general version, Theorem~\ref{general.isomorphism}, are crucial to our later results---Theorems~\ref{approximation.subgroup} and~\ref{contineous.laws}---and certain aspects of the reasoning are novel, we include a detailed proof for it in Appendix~\ref{isomorphism.proof}.

\subsection{General Isomorphism Theorem}\label{orbit.partition}
The information lattices considered in Section~\ref{coset.subsection} is rather limited---by Lagrange's theorem, coset-partitions are all equal partitions. In this subsection, we consider arbitrary information lattices---we do not require the sample space to be a group. Instead, we treat a general sample-space-partition as an \emph{orbit-partition} resulting from some group-action on the sample space.

\subsubsection{Group-Actions and Permutation Groups}
\begin{definition}
Given a group $G$ and a set $A$, a group-action of $G$ on $A$ is a function $(g,a)\mapsto g(a)$, $g\in G$, $a\in A$,  that satisfies the following two conditions:
\begin{itemize}
\item $(g_1g_2)(a)=\bigl(g_1(g_2(a)\bigr)$ for all $g_1,g_2\in G$ and $a\in A$;
\item $e(a)=a$ for all $a\in A$, where $e$ is the identity of $G$.
\end{itemize}
\end{definition}
We write $(G,A)$ to denote the group-action.

Now, we turn to the notions of \emph{orbits} and \emph{orbit-partitions}. We shall see that every group-action $(G,A)$ induces unambiguously an equivalence relation as follows. We say that $x_1$ and $x_2$ are \emph{connected} under a group-action $(G,A)$ if there exists a $g\in G$ such that $x_2=g(x_1)$. We write $x_1\conn x_2$. It is easy to check that this ``being-connected'' relation $\conn$ is an equivalence relation on $A$. By the fundamental theorem of equivalence relations, it defines a partition on $A$.

\begin{definition}
Given a group-action $(G,A)$, we call the equivalence classes with respect to the equivalence relation $\conn$, or the parts of the induced partition of $A$, the \emph{orbits} of the group-action. Accordingly, we call the induced partition the \emph{orbit-partition} of $(G,A)$ .
\end{definition}

\subsubsection{Sample-Space-Partition as Orbit-Partition}
In fact, starting with a partition $\Pi$ of a set $A$, we can go in the other direction and  unambiguously define a group action $(G,A)$ such that the orbit-partition of $(G,A)$ is exactly the given partition $\Pi$. To see this, note the following salient feature of group-actions: For any given group-action $(G,A)$, associated with every element $g$ in the group is a mapping from $A$ to itself and any such mappings must be bijective. This feature is the direct consequence of the group axioms. To see this, note that every group element $g$ has a unique inverse $g^{-1}$. According to the first defining property of group-actions, we have $(gg^{-1})(x)=g\left(g^{-1}(x)\right)=e(x)=x$ for all $x\in A$. This requires that the mappings associated with $g$ and $g^{-1}$ to be invertible. Clearly, the identity $e$ of the group corresponds to the identity map from $A$ to $A$.

With the observation that under group-action $(G,A)$ every group element corresponds to a permutation of $A$, we can treat every group as a collection of permutations that is closed under permutation composition. Specifically, for a given partition $\Pi$ of a set $A$, it is easy to check that all the permutations of $A$ that permute the elements of the parts of $\Pi$ \emph{only} to the elements of the same parts form a group. These permutations altogether form the so-called \emph{permutation representation} of $G$ (with respect to $A$). For this reason in the following, without loss of generality, we treat all groups as permutation groups. We denote by $G_{\Pi}$ the permutation group corresponding as above to a partition $\Pi$---$G_{\Pi}$ acts naturally on the set $A$ by permutation, and the orbit partition of $(G_\Pi,A)$ is exactly $\Pi$.

From group theory, we know that this orbit-partition--permutation-group-action relation is a one-to-one correspondence. Since every information element corresponds definitively to a sample-space-partition, we can identify every information element by a permutation group. Given a set $\mathbf{M}=\{m_i:i\in [n]\}$ of information elements, denote the set of the corresponding permutation groups by $\mathbf{G}=\{G_i:i\in [n]\}$. Note that all the permutations in the permutation groups $G_i$, $i\in [n]$, are permutations of the same set, namely the sample space. Hence, all the permutation groups $G_i$, $i\in [n]$, are subgroups of the symmetric group $S_{|\Omega|}$, which has order $2^{|\Omega|}$. Therefore, it makes sense to take intersection and union of groups from the collection $\mathbf{G}$.

\subsubsection{From Coset-Partition to Orbit-Partition---From Equal Partition to General Partition}
In fact, the previously studied coset-partitions are a special kind of orbit-partitions. They are orbit-partitions of group-actions defined by the native group multiplication. Specifically, given a subgroup $G_1$ of $G$, a group-action $(G_1,G)$ is defined such that $g_1(a)=g_1\circ a$ for all $g_1\in G_1$ and $a\in G$, where ``$\circ$'' denotes the native binary operation of the group $G$. The orbit-partition of such a group-action is exactly the coset-partition of the subgroup $G_1$. Therefore, by taking a different kind of group-action---permutation rather than group multiplication---we are freed from the ``equal-partition'' restriction so that we can correspond arbitrary information elements identified with arbitrary sample-space-partitions to subgroups. It turns out information lattices generated by sets of information elements and subgroup lattices generated by the corresponding sets of permutation groups remain isomorphic to each other. Thus, the isomorphism relation between information lattices and subgroup lattices holds in full generality.

\subsubsection{Isomorphism Relation Remains Between Information Lattices and Subgroup Lattices}
Similar to Section~\ref{coset.subsection}, we consider a set $\mathbf{M}=\{m_i,i\in[n]\}$ of information element. Unlike in Section~\ref{coset.subsection}, the information elements $m_i$, $i\in [n]$ considered here are arbitrary. As we discussed in the above, with each information element $m_i$ we associate a permutation group $G_i$ according to the orbit-partition--permutation-group-action correspondence. Denote the set of corresponding permutation groups by $\mathbf{G}=\{G_i,i\in [n]\}$.

\begin{theorem}{(General Isomorphism Theorem)}\label{general.isomorphism}
The information lattice $\lattice_\mathbf{M}$ is isomorphic to the subgroup lattice $\lattice_\mathbf{G}$.
\end{theorem}

The arguments for Theorem~\ref{general.isomorphism} are similar to those for Theorem~\ref{coset.isomorphism}---we demonstrate that the orbit-partition--permutation-group-action correspondence is a lattice isomorphism between $\lattice_\mathbf{M}$ and $\lattice_\mathbf{G}$.

\section{An Approximation Theorem}\label{approximation}
From this section on, we shift our focus to the quantitative aspects of the parallelism between information lattices and subgroup lattices. In the previous section, by generalizing from coset-partitions to orbit-partitions, we successfully established an isomorphism between general information lattices and subgroup lattices. In this section, we shall see that not only is the qualitative structure preserved, but also the quantitative structure---the entropy structure of information lattices---is essentially captured by their isomorphic subgroup lattices.

\subsection{Entropies of Coset-partition Information Elements}
We start with a simple and straightforward observation for the entropies of coset-partition information elements on information lattices.

\begin{proposition}\label{entropy calculation}
Let $\{G_i: i\in [n]\}$ be a set of subgroups of group $G$ and $\{m_i:i\in [n]\}$ be the set of corresponding coset-partition information elements. The entropies of the joint and common information elements on the information lattice, generated from $\{m_i:i\in [n]\}$, can be calculated from the subgroup-lattice, generated from $\{G_i: i\in [n]\}$, as follows
\begin{equation}\label{coset.entropy.joint}
h(m^{[n]})=\log \frac{|G|}{|\meet_{i\in [n]}G_i|},
\end{equation}
\textrm{ and}
\begin{equation}\label{coset.entropy.comm}
h(m_{[n]})=\log \frac{|G|}{|\join_{i\in [n]}G_i|}.
\end{equation}
\end{proposition}

Proposition~\ref{entropy calculation} follows easily from the isomorphism relation established by Theorem~\ref{general.isomorphism}.

Note that the right hand sides of both Equation~\eqref{coset.entropy.joint} and~\eqref{coset.entropy.comm} are the logarithms of the indices of subgroups. In the following, we shall call them, in short, \emph{log-indices}.

Proposition~\ref{entropy calculation} establishes a quantitative relation between the entropies of the information elements on coset-partition information lattices and the log-indices of the subgroups on the isomorphic subgroup lattices. This quantitative relation is \emph{exact}. However, the scope of Proposition~\ref{entropy calculation} is rather restrictive---it applies only to certain special kind of ``uniform'' information elements, because, by Lagrange's theorem, all coset-partitions are equal partitions.

In Section~\ref{isomorphism}, by generalizing from coset-partitions to orbit-partitions we successfully removed the ``uniformness'' restriction imposed by the coset-partition structure. At the same time, we established a new isomorphism relation, namely orbit-partition--permutation-group-action correspondence, between information lattices and subgroup lattices. It turns out that this generalization maintains an ``rough'' version of the quantitative relation established in Proposition~\ref{entropy calculation} between the entropies of information lattices and the log-indices of their isomorphic permutation-subgroup lattices. As we shall see in the next section, the entropies of the information elements on information lattices can be approximated, up to arbitrary precision, by the log-indices of the permutation groups on their isomorphic subgroup lattices.

\subsection{Subgroup Approximation Theorem}
To discuss the approximation formally, we introduce two definitions as follows.
\begin{definition}
Given an information lattice $\lattice_\mathbf{M}$ generated from a set $\mathbf{M}=\{m_i,i\in [n]\}$ of information elements, we call the real vector
$$\bigl(H(m):m\in \lattice_\mathbf{M} \bigr), $$
whose components are the entropies of the information elements on the information lattice $\lattice_\mathbf{M}$ generated by $\mathbf{M}$, listed according to a certain prescribed order, the \emph{entropy vector of $\lattice_\mathbf{M}$}, denoted  $h(\lattice_{\mathbf{M}})$.
\end{definition}

The entropy vector $h({\lattice_\mathbf{M}})$ captures the informational structure among the information elements of $\mathbf{M}$.

\begin{definition}
Given a subgroup lattice $\lattice_\mathbf{G}$ generated from a set $\mathbf{G}=\{G_i, i\in [n]\}$ of subgroups of a group $G$, we call the real vector
$$\frac{1}{|G|}\left( \log \frac{|G|}{|G'|}: G'\in \lattice_\mathbf{G} \right),$$
whose components are the normalized log-indices of the subgroups on the subgroup lattice $\lattice_\mathbf{G}$ generated by $\mathbf{G}$, listed according to a certain prescribed order, the \emph{normalized log-index vector of $\lattice_{\mathbf{G}}$}, denoted $l(\lattice_\mathbf{G})$.
\end{definition}

In the following, we assume that $l(\lattice_\mathbf{G})$ and $h(\lattice_\mathbf{M})$ are accordingly aligned.
\begin{theorem}\label{approximation.subgroup}
Let $\mathbf{M}=\{m_i,i\in [n]\}$ be a set of information elements. For any $\epsilon>0$ there exists an $N >0$ and a set $\mathbf{G}^N= \{G_i:i\in [n]\}$ of subgroups of the symmetry group $S_N$ of order $2^N$ such that
\begin{equation}
\left\Vert h(\lattice_\mathbf{M})-l(\lattice_{\mathbf{G}^N})\right\Vert<\epsilon,
\end{equation}
where ``$\left\Vert \cdot \right\Vert$'' denotes the norm of real vectors.

\end{theorem}

Theorem~\ref{approximation.subgroup} subsumes the approximation carried out by Chan and Yeung in~\cite{Chan02}, which is limited to joint entropies. The approximation procedure we carried out to prove Theorem~\ref{approximation.subgroup} is similar to that of Chan and Yeung~\cite{Chan02}---both use Stirling's approximation formula for factorials. But, with the group-action relation between information elements and permutation groups being exposed, and the isomorphism between information lattices and subgroup lattices being revealed, the approximation procedure becomes transparent and the seemingly surprising connection between information theory and group theory becomes mathematically natural. For these reasons, we included a detailed proof in Appendix~\ref{approximation.subgroup.proof}.

\section{Parallelism between Continuous Laws of Information Elements and those of Subgroups}\label{law.parallelism}
As a consequence of Theorem~\ref{approximation.subgroup}, we shall see in the following that if a continuous law holds in general for information elements, then the same law must hold for the log-indices of subgroups, and vice versa.

In the following, for reference and comparison purposes, we first review the known laws concerning the entropies of joint and common information elements. These laws, usually expressed in the form of \emph{information inequalities}, are deemed to be fundamental to information theory~\cite{Yeung02}.

\subsection{Laws for Information Elements}\label{supermodularity}

\subsubsection{Non-Negativity of Entropy}
\begin{proposition}\label{non-negative}
For any information element $m$, we have $H(m)\geq 0$.
\end{proposition}

\subsubsection{Laws for Joint Information}
\begin{proposition}\label{non-decreasing}
Given a set $\{m_i,i\in [n]\}$ of information elements, if $\alpha\subseteq \beta$, $\alpha,\beta \subseteq [n]$, then $H(m^\alpha)\leq H(m^\beta).$
\end{proposition}

\begin{proposition}\label{submodular}
For any two sets of information elements $\{m_i:i\in \alpha\}$ and $\{m_j:j\in \beta\}$, the following inequality holds:
$$H(m^{\alpha})+H(m^{\beta})\geq H(m^{\alpha\cup \beta}) + H(m^{\alpha\cap \beta}). $$
\end{proposition}
This proposition is mathematically equivalent to the following one.
\begin{proposition}\label{submodular.old}
For any three information elements $m_1$, $m_2$, and $m_3$, the following inequality holds:
$$H(m^{12})+H(m^{23})\geq H(m^{123})+H(m^{3}). $$
\end{proposition}
Note that $H(m^{3})=H(m_3)$.

Proposition~\ref{submodular} (or equivalently~\ref{submodular.old}) is usually called the submodularity law for entropy function. Proposition~\ref{non-negative},~\ref{non-decreasing},~and~\ref{submodular} are known, collectively, as the \emph{polymatroidal axioms}~\cite{Oxley92,ZhangYeung98}. Up until very recently, these are the only known laws for entropies of joint information elements.

In 1998, Zhang and Yeung discovered a new information inequality, involving four information elements \cite{ZhangYeung98}.
\begin{proposition}{(Zhang-Yeung Inequality)}\label{Zhang-Yeung}
For any four information elements $m_i$, $i=1,2,3$, and $4$, the following inequality holds:
\begin{multline}\label{Zhang-Yeung.inequality}
3H(m^{13})+3H(m^{14})+H(m^{23})+H(m^{24})+3H(m^{34})\\
\geq H(m^1)+2H(m^3)+2H(m^4)\\
\mbox{}+H(m^{12})+4H(m^{134})+H(m^{234}).
\end{multline}
\end{proposition}

This newly discovered inequality, classified as a \emph{non-Shannon type information inequality}~\cite{Yeung02}, proved that our understanding on laws governing the quantitative relations between information elements is incomplete. Recently, six more new four-variable information inequalities were discovered by Dougherty~et~al.\ \cite{Zeger06:1}.

Information inequalities such as those presented above were called ``laws of information''~\cite{Yeung02,Pippenger86}. Seeking new information inequalities is currently an active research topic~\cite{ZhangYeung98,Zhang03,Matus06,Li07:02}. In fact, they should be more accurately called ``laws of joint information'', since these inequalities involves only joint information only. We shall see below laws involving common information.

\subsubsection{Common Information v.s. Mutual Information}
In contrast to joint information, little research has been done to laws involving common information. So far, the only known non-trivial law involving both joint information and common information is stated in the following proposition, discovered by G\'{a}cs and K\"{o}rner \cite{Gacs73}.
\begin{proposition}\label{common.less}
For any two information element $m_1$ and $m_2$, the following inequality holds:
$$H(m_{12})\leq I(m_1;m_2)=H(m^{1})+H(m^2)-H(m^{12}).$$
\end{proposition}
Note that $m^1=m_1$ and $m^2=m_2$.

\subsubsection{Laws for Common Information}
Dual to the non-decreasing property of joint information, it is immediately clear that entropies of common information are non-increasing.
\begin{proposition}\label{non-increasing}
Given a set $\{m_i,i\in [n]\}$ of information elements, if $\alpha\subseteq \beta$ $\alpha,\beta\subseteq [n]$, then $H(m_\alpha)\geq H(m_\beta).$
\end{proposition}

Comparing to the case of joint information, one may naturally expect, as a dual counterpart of the submodularity law of joint information, a supermodularity law to hold for common information. In other words, we have the following conjecture.

\begin{conjecture}
For any three information elements $m_1$, $m_2$, and $m_3$, the following inequality holds:
\begin{equation}\label{supermodular}
H(m_{12})+H(m_{23})\leq H(m_{123})+H(m_{2}).
\end{equation}
\end{conjecture}

We see this conjecture as natural because of the intrinsic duality between the join and meet operations of information lattices. Due to the combinatorial nature of common information~\cite{Gacs73}, it is not obvious whether the conjecture holds. With the help of our approximation results established in Theorem~\ref{approximation.subgroup} and~\ref{contineous.laws}, we find, surprisingly, that neither the conjecture nor its converse holds. In other words, common information observes neither the submodularity nor the supermodularity law.

\subsection{Continuous Laws for Joint and Common Information}
As a consequence of Theorem~\ref{approximation.subgroup}, we shall see in the following that if a continuous law holds for information elements, then the same law must hold for the log-indices of subgroups, and vice versa. To convey this idea, we first present the simpler case involving only joint and common information elements. To state our result formally, we first introduce two definitions.

\begin{definition}
Given a set $\mathbf{M}=\{m_i:i\in [n]\}$ of information elements, consider the collection $\mathcal{M}=\{m_\alpha,m^\beta:\alpha,\beta\subseteq [n]\}$ of join and meet information elements generated from $\mathbf{M}$. We call the real vector
$$\Bigr(H(m^\alpha),H(m_\beta):\alpha,\beta \subseteq [n],\alpha, \beta\neq \Phi\Bigl),$$
whose components are the entropies of the information elements of $\mathcal{M}$, the \emph{entropy vector of $\mathcal{M}$}, denoted by $h_\mathcal{M}$.
\end{definition}

\begin{definition}
Given a set $\mathbf{G}=\{G_i:i\in [n]\}$ of subgroups of a group $G$, consider the set $\mathcal{G}=\{G_\alpha,G^\beta:\alpha,\beta\subseteq [n]\}$ of the subgroups generated from $\mathbf{G}$. We call the real vector
$$\frac{1}{|G|}\Bigl(\log\frac{|G|}{|G^\alpha|},\log\frac{|G|}{|G_\beta|}:\alpha,\beta \subseteq [n],\alpha,\beta\neq \Phi\Bigr),$$
whose components are the normalized log-indices of the subgroups in $\mathcal{M}$, the \emph{normalized log-index vector of $\mathcal{G}$}, denoted by $l_\mathcal{G}$.
\end{definition}

In this context, we assume that the components of both $l_\mathcal{G}$ and $h_\mathcal{M}$ are listed according to a common fixed order. Moreover, we note that both the vectors $h_\mathcal{M}$ and $l_\mathcal{G}$ have dimension $2^{n+1}-n-2$.

\begin{theorem}\label{contineous.laws.joint.common}
Let $f:\mathbb{R}^{2^{n+1}-n-2} \rightarrow \mathbb{R}$ be a continuous function. Then, $f(h_{\mathcal{M}})\geq 0$ holds for all sets $\mathbf{M}$ of $n$ information elements if and only if $f(l_{\mathcal{G}}) \geq 0$ holds for all sets $\mathbf{G}$ of $n$ subgroups of any group.
\end{theorem}

Theorem~\ref{contineous.laws.joint.common} is a special case of Theorem~\ref{contineous.laws}.

Theorem~\ref{contineous.laws.joint.common} and its generalization---Theorem~\ref{contineous.laws}---extend the result obtained by Chan and Yeung in~\cite{Chan02} in the following two ways. First, Theorem~\ref{contineous.laws.joint.common} and~\ref{contineous.laws} apply to all continuous laws, while only linear laws were considered in~\cite{Chan02}. Even though so far we have not yet encountered any nonlinear law for entropies, it is highly plausible that nonlinear information laws may exist given the recent discovery that at least certain part of the boundary of the entropy cones involving at least four information elements are curved~\cite{Matus2007}. Second, our theorems encompass both common information and joint information, while only joint entropies were considered in~\cite{Chan02}. For example, laws such as  Propositions~\ref{common.less} and~\ref{non-increasing} cannot even be expressed in the setting of~\cite{Chan02}. In fact, as we shall see later in Section~\ref{common.info.law}, the laws of common information depart from those of joint information very early---unlike joint information, which obeys the submodularity law, common information admits neither submodularity nor supermodularity. For these reasons, we believe that our extending the subgroup approximation to common information is of interest in its own right.

\subsection{Continuous Laws for General Lattice Information Elements}
In this section, we extend Theorem~\ref{contineous.laws.joint.common} to all the information elements in information lattices, not limited to the ``pure'' joint and common information elements. In the following, we introduce some necessary machinery to formally present the result in full generality.

Note that an element from the lattice generated from a set $X$ has its expression built from the generating elements of the lattice in the similar way that \emph{terms} are built from \emph{literals} in mathematical logic. In particular, we define \emph{lattice-terms} as follows:
\begin{definition}
An expression $E$ is called a \emph{lattice-term} formed from a set $\mathrm{X}$ of literals if either $E$ is a literal from $\mathrm{X}$ or $E$ is formed from two lattice-terms with either the \emph{join} or the \emph{meet} symbols: $E= x \, \mathrm{OP} \, y$, where $x$ and $y$ are lattice-terms and $\mathrm{OP}$ is either the join symbol $\join$ or the meet symbol $\meet$.
\end{definition}

\begin{definition}
Suppose that $E_i$, $i\in [k]$, are lattice-terms generated from a literal set of size $n$:
$\mathrm{X}=\{x_1,\cdots,x_n\}$. We call an expression of the form
$$f\bigr(H(E_1),\cdots,H(E_k)\bigl),$$
where $f$ represents a function from $\mathbb{R}^k$ to $\mathbb{R}$ and $H$ represents the entropy function, an $n$-variable \emph{generalized information expression}.
\end{definition}

We \emph{evaluate} an $n$-variable generalized information expression $f\bigl(H(E_1),\cdots,H(E_k)\bigr)$ against a set $\mathrm{M}=\{m_i:i\in [n]\}$ of information elements by substituting $x_i$ with $m_i$ respectively, calculating the entropy of the information elements obtained by evaluating the lattice-terms $E_i$ according to the semantics of the join and meet operations on information elements, and then obtaining the corresponding function value. We denote this value by
$$f\bigl(H(E_1),\cdots,H(E_k)\bigr)\bigl.\bigr|_{\mathrm{M}}.$$

\begin{definition}
If an $n$-variable generalized information expression $f\bigl(H(E_1),\cdots,H(E_k)\bigr)$ is evaluated non-negatively for any set of $n$ information elements, i.e.,
$$f\bigl(H(E_1),\cdots,H(E_k)\bigr)\bigl. \bigr|_\mathrm{M}\geq 0, \textrm{ for all } \mathrm{M},$$ then we call
$$f\bigl(H(E_1),\cdots,H(E_k)\bigr)\geq 0$$
an \emph{$n$-variable information law}.
\end{definition}

Similar to generalized information expressions, we define \emph{generalized log-index expression} as follows.
\begin{definition}
we call an expression of the form
$$f\bigr(L(E_1),\cdots,L(E_k)\bigl),$$
where $f$ represents a function from $\mathbb{R}^k$ to $\mathbb{R}$ and $L$ represents the normalized log-index function of subgroups, an $n$-variable \emph{generalized log-index expression}.
\end{definition}

We \emph{evaluate} an $n$-variable generalized log-index expression $f\bigl(L(E_1),\cdots,L(E_k)\bigr)$ against a set $\mathbf{G}=\{G_i:i\in [n]\}$ of subgroups of a group $G$ by substituting $x_i$ with $G_i$ respectively, calculating the log-index of the subgroups obtained by evaluating the lattice-terms $E_i$ according to the semantics of the join and meet operations on subgroups, and then obtaining the corresponding function value. We denote this value by
$$f\bigl(L(E_1),\cdots,L(E_k)\bigr)\bigl.\bigr|_{\mathbf{G}}.$$

\begin{definition}
If an $n$-variable generalized log-index expression $f\bigl(H(E_1),\cdots,H(E_k)\bigr)$ is evaluated non-negatively for any set of $n$ subgroups of any group, i.e.,
$$f\bigl(L(E_1),\cdots,L(E_k)\bigr)\bigl. \bigr|_\mathbf{G}\geq 0, \textrm{ for all } \mathbf{G},$$ then we call
$$f\bigl(L(E_1),\cdots,L(E_k)\bigr)\geq 0$$
an \emph{$n$-variable subgroup log-index law}.
\end{definition}

With the above formalism and corresponding notations, we are ready to state our equivalence result concerning the generalized information laws.

\begin{theorem}\label{contineous.laws}
Suppose that $f$ is continuous. Then an $n$-variable information law
$$f\bigl(H(E_1),\cdots,H(E_k)\bigr)\geq 0$$
holds if and only if the corresponding $n$-variable subgroup log-index law
$$f\bigl(L(E_1),\cdots,L(E_k)\bigr)\geq 0$$
holds.
\end{theorem}

\begin{proof}
To see one direction, namely that $f\bigr(L(E_1),\cdots,L(E_k)\bigl)\geq 0$ implies that $f\bigr(H(E_1),\cdots,H(E_k)\bigl)\geq 0$, assume that there exists a set $\mathbf{M}$ of information elements such that $f\bigr(H(E_1),\cdots,H(E_k)\bigl)\bigl. \bigr|_{\mathbf{M}}=a$ for some $a<0$. By the continuity of the function $f$ and Theorem~~\ref{approximation.subgroup}, we are guaranteed to be able to construct, from the information lattice generated from $\mathbf{M}$, some subgroup lattice $\lattice_\mathbf{G}$ such that the value of the function $f$ at the normalized log-indices of the correspondingly constructed subgroups is arbitrarily close to $a<0$. This contradicts the assumption that $f\bigl(L(E_1),\cdots,L(E_k)\bigr)\bigl. \bigr|_\mathbf{G} \geq 0$ holds for all sets $\mathbf{G}$ of $n$ subgroups of any group.

On the other hand, for any normalized log-indices of the subgroups from subgroup lattices, it can be readily interpreted as the entropies of information elements by taking permutation representation for the subgroups on the subgroup lattice and then producing an information lattice, according to the orbit-partition--permutation-group-action correspondence. Therefore, that $f\bigl(H(E_1),\cdots,H(E_k)\bigr)\bigl. \bigr|_\mathbf{M}\geq 0$  holds for all sets $\mathbf{M}$ implies that $f\bigl(L(E_1),\cdots,L(E_k)\bigr)\bigl. \bigl|_{\mathbf{G}}\geq 0$ holds for all sets $\mathbf{G}$.
\end{proof}

\subsection{Common Information Observes Neither Submodularity Nor Supermodularity Laws}\label{common.info.law}
As discussed in the above, appealing to the duality between the join and the meet operations, one might conjecture, dual to the well-known submodularity of joint information, that common information would observe the supermodularity law. It turns out that common information observes neither the submodularity~\eqref{submodularity.invalid} nor the supermodularity~\eqref{supmodularity.invalid} law---neither of the following two inequalities holds in general:

\begin{eqnarray}
h(m_{12})+h(m_{23})\geq h(m_{123})+h(m_2) \label{submodularity.invalid}\\
h(m_{12})+h(m_{23})\leq h(m_{123})+h(m_2). \label{supmodularity.invalid}
\end{eqnarray}

Because common information is combinatorial in flavor---it depends on the ``zero pattern'' of joint probability matrices~\cite{Gacs73}---it is hard to directly verify the validity of~\eqref{submodularity.invalid} and~\eqref{supmodularity.invalid}. However, thanks to Theorem~\ref{contineous.laws}, we are able to construct subgroup counterexamples to invalidate~\eqref{submodularity.invalid} and~\eqref{supmodularity.invalid} indirectly.

To show that~\eqref{supmodularity.invalid} fails, it suffices to find three subgroups $G_1, G_2$, and $G_3$ such that
\begin{equation}\label{counter.supmodularity.group}
|G_1\join G_2||G_2\join G_3| < |G_1\join G_2 \join G_3| |G_2|.
\end{equation}
Consider $G=S_5$, the symmetry group of order $2^5$, and its subgroups $G_1=\gen{(12345)}$, $G_2=\gen{(12)(45)}$, and $G_3=\gen{(12543)}$. The subgroup $G_1$ is the permutation group generated by permutation $(12345)$, $G_2$ by $(12)(45)$, and $G_3$ by $(12543)$. (Here, we use the standard cycle notation to represent permutations.) Consequently, we have $G_1\join G_2=\gen{(12345),(12)(45)}$, $G_2\join G_3=\gen{(12543),(12)(45)}$, and $G_1\join G_2 \join G_3=\gen{(12345),(12)(45),(12543)}$. It is easy to see that both $G_1\join G_2$ and $G_2 \join G_3$ are dihedral groups of order 10 and that $G_1\join G_2\join G_3$ is the alternative group $A_5$, hence of order $60$. The order of $G_2$ is 2. Therefore, we see that the subgroups $G_1$, $G_2$, and $G_3$ satisfy~\eqref{counter.supmodularity.group}. By Theorem~\ref{contineous.laws}, the supermodularity law~\eqref{supmodularity.invalid} does not hold in general for common information. (Thank to Professor Eric Moorhouse for contributing this counterexample.)

Similar to the case of supermodularity, the example with $G_2=\{e\}$ and $G_1=G_3=G$, $|G|\neq 1$, invalidates the group version of~\eqref{submodularity.invalid}. Therefore, according to Theorem~\ref{contineous.laws}, the submodularity law~\eqref{submodularity.invalid} does not hold in general for common information either.

\section{Discussion}\label{discussion}
This paper builds on some of Shannon's little-recognized legacy and adopts his interesting concepts of information elements and information lattices. We formalize all these concepts and clarify the relations between random variables and information elements, information elements and $\sigma$-algebras, and, especially, the one-to-one correspondence between information elements and sample-space-partitions. We emphasize that such formalization is conceptually significant. As demonstrated in this paper, beneficial to the formalization carried out, we are able to establish a comprehensive parallelism between information lattices and subgroup lattices. This parallelism is mathematically natural and admits intuitive group-action explanations. It reveals an intimate connection, both structural and quantitative, between information theory and group theory. This suggests that group theory might serve a promising role as a suitable mathematical language in studying deep laws governing information.

Network information theory in general, and capacity problems for network coding specifically, depend crucially on our understanding of intricate structures among multiple information elements. By building a bridge from information theory to group theory, we can now access the set of well-developed tools from group theory. These tools can be brought to bear on certain formidable problems in areas such as network information theory and network coding. Along these lines, by constructing subgroup counterexamples we show that neither the submodularity nor the supermodularity law holds for common information, neither of which is obvious from traditional information theoretic perspectives.

\appendices
\section{Proof of Theorem~\ref{coset.isomorphism}}\label{isomorphism.proof}
\begin{proof}
To show two lattices are isomorphic, we need to demonstrate a mapping, from one lattice to the other, such that it is a lattice-morphism---it honors both join and meet operations---and bijective as well. Instead of proving that $\lattice_\mathbf{G}$ is isomorphic to $\lattice_G$ directly, we show that the dual of $\lattice_\mathbf{G}$ is isomorphic to $\lattice_\mathbf{M}$. Figuratively speaking, the dual of a lattice $\lattice$ is the lattice obtained by flipping $\lattice$ upside down. Formally, the dual lattice $\lattice'$ of a lattice $\lattice$ is the lattice defined on the same set with the partial order reversed. Accordingly, the join operation of the prime lattice $\lattice$ corresponds to the meet operation for the dual lattice $\lattice'$ and the meet operation of $\lattice$ to the join operation for $\lattice'$. In the other words, we show that $\lattice_G$ is isomorphic to $\lattice_\mathbf{M}$ by demonstrating a bijective mapping $\phi: \lattice_\mathbf{G} \rightarrow \lattice_\mathbf{M}$ such that
\begin{equation}\label{joint.info}
\phi(G\join G') = \phi(G)\meet \phi(G'),
\end{equation}
and
\begin{equation}\label{comm.info}
\phi(G\meet G') = \phi(G)\join \phi(G'),
\end{equation}
hold for all $G,G'\in \lattice_\mathbf{G}$.

Note that each subgroups on the subgroup lattice $\lattice_\mathbf{G}$ is obtained from the set $\mathbf{G}=\{G_i:i\in[n]\}$ via a sequence of join and meet operations and each information element on the information lattice $\lattice_\mathbf{M}$ is obtained similarly from the set $\mathbf{M}=\{m_i:i\in [n]\}$. Therefore, to show that $\lattice_\mathbf{G}$ is isomorphic to $\lattice_\mathbf{M}$, according to the induction principle, it is enough to demonstrate a bijective mapping $\phi$ such that
\begin{itemize}
\item $\phi(G_i)=m_i$, for all $G_i\in \mathbf{G}$ and $m_i\in \mathbf{M}$;
\item For any $G,G'\in \lattice_\mathbf{G}$, if $\phi(G)=m$ and $\phi(G')=m'$, then
\begin{equation}\label{joint.info.induction}
\phi(G\join G')=m\meet m', \text{ and }
\end{equation}
\begin{equation}\label{comm.info.induction}
 \phi(G\meet G')=m\join m'.
\end{equation}
\end{itemize}

Naturally, we take $\phi: \lattice_\mathbf{G} \rightarrow \lattice_\mathbf{M}$ to be the mapping that assigns to each subgroup $G\in \lattice_\mathbf{G}$ the information element identified by the coset-partition of the subgroup $G$. Thus, the initial step of the induction holds by assumption. On the other hand, it is easy to see that the mapping $\phi$ so defined is bijective simply because different subgroups always produce different coset-partitions and vice versa. Therefore, we are left to show that Equation~\eqref{joint.info.induction} and~\eqref{comm.info.induction} holds.

We first show that $\phi$ satisfies Equation~\eqref{joint.info.induction}. In other words, we show that the coset-partition of the intersection subgroup $G\cap G'$ is the coarsest among all the sample-space-partitions that are finer than both the coset-partitions of $G$ and $G'$. To see this, let $\Pi$ be a sample-space-partition that is finer than both the coset-partitions of $G$ and $G'$ and $\pi$ be a part of $\Pi$. Since $\Pi$ is finer than the coset-partitions of $G$, $\pi$ must be contained in some coset $C$ of $G$. For the same reason, $\pi$ must be contained in some coset $C'$ of $G'$ as well. Consequently, $\pi\subseteq C'\cap C'$ hold. Realizing that $C\cap C'$ is a coset of $G\cap G'$, we conclude that the coset-partition of $G\cap G'$ is coarser than $\Pi$. Since $\Pi$ is chosen arbitrary, this proves that the coset-partition of the intersection subgroup $G\cap G'$ is the coarsest among all the sample space partitions that are finer than both the coset-partitions of $G$ and $G'$. Therefore, Equation~\eqref{joint.info.induction} holds for $\phi$.

The proof for Equation~\eqref{comm.info.induction} is more complicated. We use an idea called ``transitive closure''. Similarly, we need to show that the coset-partition of the subgroup $G\join G'$ generated from the union of $G$ and $G'$ is the finest among all the sample-space-partitions that are coarser than both the coset-partitions of $G$ and $G'$. Let $\Pi$ be a sample-space-partition that is coarser than both the coset-partitions of $G$ and $G'$. Denote the coset partition of the subgroup $G\join G'$ by $\bar{Pi}$. Let $\bar{\pi}$ be a part of $\bar{\Pi}$. It suffices to show that $\bar{\pi}$ is contained in some part of $\Pi$. Pick an element $x$ from $\bar{\pi}$. This element $x$ must belong to some part $\pi$ of $\Pi$. It remains to show $\bar{\pi} \subseteq \pi$. In other words, we need to show that $y\in \pi$ for any $y\neq x, y\in\pi^{ij}$.  Note that $\pi$ is a part of the coset-partition of the subgroup $G_i\join G_j$. In other words, $\pi$ is a coset of $G_i\join G_j$. The following reasoning depends on the following fact from group theory~\cite{Dummit99}.
\begin{proposition}\label{coset.characterization}
Two elements $g_1$ and $g_2$ belong to a same (right) coset of a subgroup if and only if $g_1g_2^{-1}$ belongs to the subgroup.
\end{proposition}

Since $x$ and $y$ belong to a same coset $\pi$ of the subgroup $G\join G'$, we have $yx^{-1}\in G\join G'$. Note that any element $g$ from $G\join G'$ can be written in the form of $g=a_1b_1a_1b_2\cdots a_K b_K$ where $a_k\in G$ and $b_k\in G'$ for all $k\in [K]$. Suppose $yx^{-1}=g=a_1b_1a_1b_2\cdots a_K b_K$. We have
$$y=a_1b_1a_2b_2\cdots a_K b_K x.$$
In the following we shall show that $y$ belongs to $\bar{\pi}$ by induction on the sequence $a_1b_1\cdots a_K b_K$.

First, we claim $b_K x\in \bar{\pi}$. To see this, note that $x\in \bar{\pi}$. Since $(b_K x) x^{-1}=b_K\in G'$, by Proposition~\ref{coset.characterization}, we know that $b_K x$ and $x$ belong to a same coset $C_K$ of $G'$. By assumption, the partition $\Pi$ is coarser than the coset-partition of $G'$, the coset $C_K$ must be contained in $\bar{\pi}$, since it already contains an element $x$ of $C_K$.

For the same reason, with $b_K x \in \bar{\pi}$ showed, we can see that $a_K b_K x$ belongs to $\bar{\pi}$ as well, because $(a_K b_K x) (b_K x)^{-1}=a_K\in G$ implies $a_K b_K x$ and $b_K x$ belong to a same coset of $G$.

Continuing the above argument inductively on the sequence $a_1b_1\cdots a_K b_K$, we can finally have $a_1b_1 \cdots a_K b_K x \in \bar{\pi}$. Therefore, we have $y\in \bar{\pi}$. This concludes the proof.
\end{proof}

\section{Proof of Theorem~\ref{approximation.subgroup}}\label{approximation.subgroup.proof}
\begin{proof}
The approximation process is decomposed into three steps. The first step is to ``dilate'' the sample space such that we can turn a non-uniform probability space into a uniform probability space. The sample space partitions of the information elements are accordingly ``dilated'' as well. After dilating the sample space, depending on the approximation error tolerance, i.e., $\epsilon$, we may need to further ``amplify'' the sample space. Then, we follow the same procedure as in Section~\ref{orbit.partition} and construct a subgroup lattice using the orbit-partition--permutation-group-action correspondence.

We assume the probability measure $\mathbf{P}$ on the sample space are rational. In other words, the probabilities of the elementary event $\p_i=\Pr\{\omega_i\}$, $\omega_i\in \Omega$ are all rational numbers, namely $\p_i=\frac{p_i}{q_i}$ for some $p_i,q_i\in \mathbb{N}$.  This assumption is reasonable, because any finite dimensional real vector can be approximated, up to an arbitrary precision, by some rational vector.

Let $M$ be the least common multiple of the set $\{q_i\}$ of denominators. We ``split'' each sample point in $\Omega$ into $\frac{M p_i}{q_i}$ points. Note that $\frac{M p_i}{q_i}$ is integral.  We need to  accordingly ``dilate'' the sample space partitions of the information elements. Specifically, for each part $\pi$ of the partition of every information element $m_i$, its ``dilated'' partition $\pi'$, in the dilated sample space $\hat{\Omega}$,  contains exactly all the sample points that are ``split'' from the sample points in $\pi$. The dilated sample space $\hat{\Omega}$ has size of $\sum_{\omega_i\in \Omega} \frac{M p_i}{q_i} $. To maintain the probability structure, we assign to each sample point in the dilated sample space $\hat{\Omega}$ probability $\frac{1}{|\hat{\Omega}|}$. In other words, we equip the dilated sample space with a uniform probability measure. It is easy to check that the entire (quantitative) probability structure remains the same. Thus, we can consider all the information elements as if defined on the dilated probability space.

If necessary, depending on the approximation error tolerance $\epsilon$, we may further ``amplify'' the dilated sample space $\hat{\Omega}$ by $K$ times by ``splitting'' each of its sample points into to $K$ points. At the same time, we scale the probability of each sample point in the post-amplification sample space down by $K$ times to $\frac{1}{K|\hat{\Omega}|}$. By abusing of notation, we still use $\hat{\Omega}$ to denote the post-amplification sample space. Similar to the ``dilating'' process, all the partitions are accordingly amplified.

Before we move to the third step, we compute entropies for information elements in terms of the cardinality of the parts of its dilated sample space partition. Consider an information element $m_i$. Denote its pre-dilation sample space partition by $\Pi_i=\{\pi^j_i, j\in [J]\}$ and its post-amplification sample space partition by $\hat{\Pi}_i=\{\hat{\pi}^j_i, j\in [J]\}$. It is easy to see that the entropy $H(m_i)$ can be calculated as follows:
\begin{equation}\label{entropy.partition}
\begin{split}
H(m_i)&=-\sum_{j\in [J]} \Pr\{\pi_i^j\} \log \Pr\{\pi_i^j\}\\
&= -\sum_{j\in [J]} \Pr\{\hat{\pi}_i^j\} \log \Pr\{\hat{\pi}_i^j\}\\
&=-\sum_{j\in [J]} \frac{|\hat{\pi}^i_j|}{|\hat{\Omega}|} \log \frac{|\hat{\pi}^i_j|}{|\hat{\Omega}|}.
\end{split}
\end{equation}
All the entropies of the other information elements, including the joint and common information elements, on the entire information lattices can be computed in the exactly same way in terms of the cardinalities of the parts of their dilated sample space partitions.

In the third step, we follow the same procedure as in Section~\ref{orbit.partition},  and construct, based on the orbit-partition--permutation-group-action correspondence, a subgroup lattice that isomorphic to the information lattice generated by the set of information elements $\{m_i: i\in [n]\}$. More specifically, the subgroup lattice is constructed according to their ``post-amplification'' sample space partitions.

Suppose, on the constructed subgroup lattice, the permutation groups $G_i$ corresponds to the information element $m_i$. As in the above, the ``post-amplification'' sample space partition of $m_i$ is $\hat{\Pi}_i=\{\hat{\pi}^j_i, j\in [J]\}$. Then, the cardinality of the permutation group is simply $$|G_i|=\prod_{j\in J} \hat{\pi}_i^j!.$$
According to the isomorphism relation established in Theorem~\ref{general.isomorphism}, the above calculations remain valid for all the subgroups on the subgroup lattices.

Recall that all the groups on the subgroup lattice are permutation groups and are all subgroups of the symmetry group of order $|\hat{\Omega}|$. So the log-index of $G_i$, corresponding to $m_i$, is
\begin{equation}\label{log.index}
\log \frac{|\hat{\Omega}|!}{|G_i|}= \log  \frac{|\hat{\Omega}|!}{\prod_{j\in J} \hat{\pi}_i^j!}.
\end{equation}

As we see from Equation~\eqref{coset.entropy.joint} and \eqref{coset.entropy.comm} of Proposition~\ref{entropy calculation}, the entropies of the coset-partition information elements on information lattices equal \emph{exactly} the log-indices of their subgroups on subgroup lattices. However, for the information lattice generated from general information elements, namely information elements with non-equal sample space partitions, as we see from Equation~\eqref{entropy.partition} and \eqref{log.index}, the entropies of the information elements on the information lattice does not equal the log-indices of their corresponding permutation groups on the subgroup lattices exactly any more. But, as we can shall see, the entropies of the information elements are well \emph{approximated} by the log-indices of their corresponding permutation groups. Recall the following Stirling's approximation formula for factorials:
\begin{equation}\label{Stirling}
\log n!= n\log n -n +  o(n).
\end{equation}
``Normalizing'' the log-index in Equation~\eqref{log.index} by a factor $\frac{1}{|\hat{\Omega}|}$ and then substituting the factorials with the above Stirling approximation formula, we get
\begin{equation*}
\begin{split}
\frac{1}{|\hat{\Omega}|} \log \frac{|\hat{\Omega}|}{|G_i|}&=\frac{1}{|\hat{\Omega}|}\Bigl( |\hat{\Omega}|\log |\hat{\Omega}| -|\hat{\Omega}|-\Bigr.\\
 &\Bigl. \bigl(\sum_{j\in [J]} |\hat{\pi}_i^j|\log |\hat{\pi}_i^j| -|\hat{\pi}_i^j|\bigr) +o(|\hat{\Omega}|)\Bigr).
\end{split}
\end{equation*}
Note that in the above substitution process, we combined some finite $o(|\hat{\Omega|})$ terms ``into'' one $o(|\hat{\Omega|})$ term.

It is clear that $\sum_{j\in [J]} |\hat{\pi}_i^j|=|\hat{\Omega}|$, since $\{\hat{\pi}^j_i: j\in[J]\}$ forms a partition of $\hat{\Omega}$. Therefore, we get
\begin{equation*}
\begin{split}
\frac{1}{|\hat{\Omega}|} \log \frac{|\hat{\Omega}|}{|G_i|}&=\frac{1}{|\hat{\Omega}|}\bigl( |\hat{\Omega}|\log |\hat{\Omega}|- \sum_{j\in [J]} |\hat{\pi}_i^j|\log |\hat{\pi}_i^j| +o(\hat{\Omega})\bigr)\\
&=h(m_i)+\frac{o(|\hat{\Omega}|)}{|\hat{\Omega}|}.
\end{split}
\end{equation*}
So, the difference between the entropy $H(m_i)$ and the normalized log-index of its corresponding permutation subgroup $G_i$ diminishes for $\hat{\Omega}$ large.

Since both the entropy vector $h_\mathbf{M}$ and the log-index vector $l_{\mathbf{G}^N}$ are of finite dimension, it follows easily
$$\left\Vert h_{\mathbf{M}}-\frac{l_{\mathbf{G}^N}}{N}\right \Vert_1 =\frac{o(|\hat{\Omega}|)}{|\hat{\Omega}|}\rightarrow 0,$$
with
$$N=|\hat{\Omega}|=K \sum_{\omega_i\in \Omega} \frac{Mp_i}{q_i} \rightarrow \infty, \textrm{ by taking } K \rightarrow \infty.$$

This concludes the proof.
\end{proof}

\IEEEtriggeratref{31}

\bibliographystyle{IEEEtran}
\bibliography{research}

\end{document}